\documentclass[showpacs,aps]{revtex4}

\usepackage{graphicx}
\usepackage{amsmath}

\begin{document}

\title{Complete hyperentangled-Bell-state analysis for photon systems assisted by  quantum-dot spins in optical microcavities\footnote{Published in Optics Express \textbf{20}, 24664-24677  (2012)}}

\author{Bao-Cang Ren,  Hai-Rui Wei,  Ming Hua,  Tao Li, and Fu-Guo Deng\footnote{Corresponding author. Email address:
fgdeng@bnu.edu.cn} }
\address{Department of Physics, Applied Optics Beijing Area Major Laboratory, Beijing Normal University, Beijing 100875, China }
\date{\today }

\begin{abstract}
Bell-state analysis (BSA) is essential in quantum communication, but
it is impossible to distinguish unambiguously the four Bell states
in the polarization degree of freedom (DOF) of two-photon systems
with only linear optical elements, except for the case in which the
BSA is assisted with  hyperentangled states, the simultaneous
entanglement in more than one DOF. Here, we propose a scheme to
distinguish  completely the 16 hyperentangled Bell states in both
the polarization and the spatial-mode DOFs of two-photon systems, by
using the giant nonlinear optics in quantum dot-cavity systems. This
scheme can be applied  to increase the channel capacity of
long-distance quantum communication based on hyperentanglement, such
as entanglement swapping, teleportation, and  superdense coding. We
use hyperentanglement swapping as an example to show the application
of this HBSA.
\end{abstract}


\maketitle


\section{Introduction}

Entanglement is a key quantum resource for quantum information
processing and it plays a critical role in many important
applications in quantum communication, such as quantum key
distribution \cite{QKD1,QKD2,QKD3,QKD4,QKD5}, quantum dense coding
\cite{DC1,DC2}, quantum teleportation \cite{QT1}, and entanglement
swapping \cite{QS1}. Some important goals in quantum communication
require the complete and deterministic analysis of the Bell states.
In 1999, Vaidman's \cite{BSA1} and L$\ddot{u}$tkenhau's \cite{BSA2}
groups put forward a Bell-state analysis (BSA) for teleportation
with only linear optical elements. Unfortunately, with
linear-optical elements, one can obtain the optimal success
probability of 50\% both in theory \cite{BSAT} and in experiment
\cite{BSAE1,BSAE2,BSAE3} (Of course, by guessing at random when an
ambiguous result is obtained, one can technically achieve a 75\%
success rate in identifying the four Bell states).  The BSA on
photon pairs entangled in one degree of freedom (DOF) attracted much
attention
\cite{polar1,polar2,spa-exp,orbital,trans,time,bin,sim,2photon1,2photon2}.

The entanglement of photon pairs in several DOFs
\cite{heper1,heper2,heper3}, called it hyperentanglement, is useful
in quantum information processing, especially in quantum
communication for completing BSA on polarizations of photon pairs
\cite{BSA1kwiat,BSA2walborn,BSA3, BSA4, BSA5},  performing
entanglement purification
\cite{EPPsimon,EPPsheng1,EPPsheng2,EPPsheng3,EPPlixh,EPPdeng},
distributing entangled polarization states faithfully
\cite{shengdistribution}, or improving the channel capacity. For
example, Kwiat and Weinfurter \cite{BSA1kwiat} first introduced the
way to distinguish the four orthogonal Bell states of photon pairs
in the polarization DOF with the hyperentanglement in both the
polarization DOF and the momentum DOF in 1998. In 2003, Walborn et
al. \cite{BSA2walborn} proposed a simple scheme for completing
Bell-state measurement for photon pairs entangled in the
polarization DOF or the momentum DOF by using hyperentangled states
with linear optics. The experiments of a complete BSA have also been
reported with polarization-time-bin hyperentanglement \cite{BSA3}
and polarization-momentum hyperentanglement \cite{BSA4} later.  It
implies a complete BSA in the polarization DOF can be accomplished
with hyperentanglement in a larger Hilbert space by introducing
other DOFs. In 2008, Barreiro et al. \cite{BSA5} beat the channel
capacity limit for linear photonic superdense coding with
polarization-orbital-angular-momentum hyperentanglement. In 2002,
Simon and Pan \cite{EPPsimon} proposed an entanglement purification
protocol (EPP) using hyperentanglement in both the polarization and
the spatial DOFs. In 2008,  an efficient EPP based on a parametric
down-conversion source was proposed, resorting to this
hyperentanglement \cite{EPPsheng1}. In 2010, deterministic EPPs were
proposed with hyperentanglement
\cite{EPPsheng2,EPPsheng3,EPPlixh,EPPdeng}. In 2010, a faithful
entanglement distribution scheme for polarization  was proposed
\cite{shengdistribution}, resorting to the stability of the
frequency entanglement of photon pairs.

Considering a large Hilbert space with an additional DOF, e.g.,  a
quantum system in a hyperentangled state in two DOFs which span the
Hilbert space with 16 orthogonal Bell states, one can not
distinguish them completely with only linear optics. In 2007, Wei et
al. \cite{HBSA} pointed out that 7 states in the group of 16
orthogonal Bell states is distinguishable with only linear optics.
In 2011,  Pisenti et al. \cite{HBSA1} presented a very general
theoretical explanation of the inadequacy of linear evolution and
local measurement in (hyperentangled-) Bell-state analysis, and they
pointed out the limitations for manipulation and measurement of
entangled systems with inherently linear, unentangling devices. If
nonlinear optics is introduced, these 16 orthogonal Bell states can
be distinguished completely. In 2010, Sheng et al. \cite{kerr}
presented a complete hyperentangled BSA (HBSA) with cross-Kerr
nonlinearity. Although a lot of works have been studied on
cross-Kerr nonlinearity \cite{kerr1}, a clean cross-Kerr
nonlinearity in the optical single-photon regime is still quite a
controversial assumption with current technology \cite{kerr2,kerr3}.
In recent years, a solid state system based on an electron spin in a
quantum dot (QD) has attracted much attention with its giant
nonlinearity. In 2008, Hu et al. \cite{QD1} proposed a quantum
nondemolition method using the interaction of left-circularly and
right-circularly polarized lights with a one-side QD-cavity system.
This nonlinear optics in a  QD-cavity system can be used to
construct multi-photon entangler \cite{QD1,QD2} and photonic
polarization BSA \cite{QD3,QD4}. In 2010, Bonato et al. \cite{QD3}
proposed a photonic polarization BSA using quantum-dot microcavities
in the weak-coupling limit. In 2011, Hu et al. \cite{QD4} presented
some interesting schemes for BSA using the nonlinear optics of a
single quantum-dot spin in a one-side optical microcavity.

In this article, we present complete  HBSA  with the nonlinear
optics based on a one-side QD-cavity system. It can be used to
distinguish completely the 16 hyperentangled Bell states in both the
polarization and the spatial-mode DOFs of two-photon systems. This
scheme divides the process for HBSA into two steps. The first step
is to distinguish the four Bell states in spatial-mode DOF, without
destroying the two-photon system itself and its state in the
polarization DOF. This task should resort to quantum nondemolition
detectors (QNDs) based on nonlinear optics of a one-side QD-cavity
system. In the second step, one can adjust the QD-cavity systems to
distinguish the four Bell states in the polarization DOF. This HBSA
scheme can be applied to increase the channel capacity of
long-distance quantum communication based on hyperentanglement, such
as  entanglement swapping, teleportation, and superdense coding. We
use hyperentanglement swapping as an example to show its
application.

\section{Interaction between a circularly polarized light and a QD-cavity
system}

Considering a singly charged QD in a cavity, e.g., a self-assembled
In(Ga)As QD or a GaAs interface QD inside an optical resonant
microcavity, the QD is located in the center of the cavity to
achieve a maximal light-matter coupling. With an excess electron
injected into the QD, the singly charged QD shows the optical
resonance with the negatively charged exciton  $X^-$ that consists
of two electrons bound to one hole \cite{QD5}. According to Pauli's
exclusion principle, $X^-$ has spin-dependent transitions
\cite{QD6}. If the excess electron in the QD is in the spin state
$|\uparrow\rangle$, only the left circularly polarized light
$|L\rangle$ can be resonantly absorbed to create the negatively
charged exciton in the state $|\uparrow\downarrow\Uparrow\rangle$
with  two  antiparallel electron spins. Here $|\Uparrow\rangle$
represents a heavy-hole spin state $|+\frac{3}{2}\rangle$. If the
excess electron in the QD is in the spin state $|\downarrow\rangle$,
only the right circularly polarized light $|R\rangle$ can be
resonantly absorbed to create the negatively charged exciton in the
state $|\downarrow\uparrow\Downarrow\rangle$ with two antiparallel
electron spins. Here $|\Downarrow\rangle$ represents the heavy-hole
spin state $|-\frac{3}{2}\rangle$. They  have different phase shifts
when the photons in these two different circularly polarized states
are reflected from the QD-cavity system.

\begin{figure}[htbp]                 
\centering\includegraphics[width=8cm]{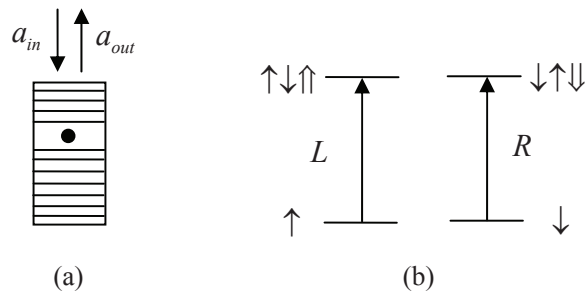} \caption{The
spin-dependent transitions for negatively charged exciton  $X^-$.
(a) A charged QD inside a micropillar microcavity with circular
cross section. (b) Spin selection rule for optical transitions of
negatively charged exciton $X^-$ due to the Pauli's exclusion
principle. $L$ and $R$ represent the left and the right circularly
polarized lights, respectively.} \label{figure1}
\end{figure}

The whole process can be represented by Heisenberg equations for the
cavity field operator $a$ and $X^-$ dipole operator $\sigma_-$ in
the interaction picture \cite{QD7},
\begin{equation}                           \label{eq.1}
\begin{split}
\frac{d a}{dt} \;\; &= \;\; -[i(\omega_c-\omega) + \frac{\kappa}{2} + \frac{\kappa_s}{2}] a -g\sigma_--\sqrt{\kappa}\, a_{in},  \\
\frac{d\sigma_-}{dt} \;\; &= \;\; -[i(\omega_{X^-}- \omega) +
\frac{\gamma}{2}] \sigma_- - g\sigma_z \, a,\\
a_{out} \;\; &= \;\;  a_{in}+\sqrt{\kappa}\, a,
 \end{split}
\end{equation}
where $\omega$, $\omega_c$, and $\omega_{X^-}$ are the frequencies
of the input probe light, cavity mode, and $X^-$ transition,
respectively. $g$ is the coupling strength between $X^-$ and the
cavity mode, $\gamma/2$ and $\kappa/2$  are the decay rates of $X^-$
and the cavity field, and $\kappa_s/2$ is the side leakage rate of
the cavity.

With a weak excitation condition ($X^-$ stays in the ground state at
most time and $\langle\sigma_z\rangle=-1$), the reflection
coefficient for the QD-cavity system can be obtained as  \cite{QD1}
\begin{eqnarray}                           \label{eq.2}
r(\omega) \;=\;
1-\frac{\kappa[i(\omega_{X^-}-\omega)+\frac{\gamma}{2}]}{[i(\omega_{X^-}-\omega)
+\frac{\gamma}{2}][i(\omega_c-\omega)+\frac{\kappa}{2}+\frac{\kappa_s}{2}]+g^2}.
\end{eqnarray}
One can get the reflection coefficient  $r_0(\omega)$ for a cold
cavity with the uncoupled QD by taking $g=0$ as following
\cite{QD1}:
\begin{eqnarray}                           \label{eq.3}
r_0(\omega) \; = \;
\frac{i(\omega_c-\omega)-\frac{\kappa}{2}+\frac{\kappa_s}{2}}{i(\omega_c-\omega)+\frac{\kappa}{2}+\frac{\kappa_s}{2}}.
\end{eqnarray}
If the excess electron is in the spin state $|\uparrow\rangle$, the
$|L\rangle$ light feels a hot cavity (coupled with the QD-cavity
system) and gets a phase shift of $\varphi_h$ after being reflected
(the subscript $h$ represents a hot cavity), whereas the $|R\rangle$
light feels a cold cavity and gets a phase shift of $\varphi_0$. By
adjusting the frequencies $\omega$ and $\omega_c$, one can get  the
reflection coefficients $|r_0(\omega)|\cong1$  for a cold cavity and
$|r_h(\omega)|\cong1$ for  a hot cavity. As the linearly polarized
probe beam can be regarded as the superposition of two circularly
polarized components $(|R\rangle+|L\rangle)/\sqrt{2}$, the state of
the reflected light becomes
$(e^{i\varphi_0}|R\rangle+e^{i\varphi_h}|L\rangle)/\sqrt{2}$ after
it is reflected from the one-side QD-cavity system. Conversely, if
the excess electron is in the spin state $|\downarrow\rangle$, the
$|L\rangle$ light feels a cold cavity and gets a phase shift of
$\varphi_0$ after being reflected, while the $|R\rangle$  light
feels a hot cavity and gets a phase shift of $\varphi_h$. The linear
polarized probe beam $(|R\rangle+|L\rangle)/\sqrt{2}$ becomes
$(e^{i\varphi_h}|R\rangle+e^{i\varphi_0}|L\rangle)/\sqrt{2}$ after
being reflected. The polarization direction of the reflected light
rotates an angle
$\theta_F^\uparrow=(\varphi_0-\varphi_h)/2=-\theta_F^\downarrow$,
which is the so-called Faraday rotation.

If the electron is in a superposition spin state
$|\psi\rangle=(|\uparrow\rangle+|\downarrow\rangle)/\sqrt{2}$ and
the photon is in the state $(|R\rangle + |L\rangle)/\sqrt{2}$, after
being reflected, the light-spin state evolves as
\begin{eqnarray}                           \label{eq.4}
\frac{1}{2}(|R\rangle+|L\rangle)\otimes(|\uparrow\rangle+|\downarrow\rangle)
&\rightarrow & \frac{1}{2}e^{i\varphi_0}[(|R\rangle
+e^{i\Delta\varphi}|L\rangle) |\uparrow\rangle +
(e^{i\Delta\varphi}|R\rangle+|L\rangle)|\downarrow\rangle],
\end{eqnarray}
where $\Delta\varphi=\varphi_h-\varphi_0$,
$\varphi_0=arg[r_0(\omega)]$, and $\varphi_h=arg[r_h(\omega)]$. In a
one-side cavity, due to spin selection rule above,  $|L\rangle$ and
$|R\rangle$ lights pick up two different phase shifts after being
reflected from  the QD-cavity system, and then the state of the
system composed of the light and the excess electron becomes an
entangled one.

\section{Complete HBSA using one-side QD-cavity systems}

A hyperentangled two-photon Bell state in both the polarization and
the spatial-mode DOFs has the  form as
\begin{eqnarray}                           \label{eq.5}
|\Phi^+\rangle^{AB}_{PS} \;\; = \;\;
\frac{1}{2}(|RR\rangle+|LL\rangle)^{AB}_P\otimes(|a_1b_1\rangle+|a_2b_2\rangle)^{AB}_S.
\end{eqnarray}
Here, the superscripts $A$  and $B$ represent the two photons in the
hyperentangled state. The subscript $P$ denotes the polarization DOF
and $S$ is the spatial-mode DOF. $a_1$ ($b_1$) and $a_2$ ($b_2$) are
the different spatial modes for the photon  $A$ ($B$). We denote the
four Bell states in the polarization DOF as
\begin{equation}  \label{eq.6}
\begin{split}
|\phi^\pm\rangle^{AB}_{P} \;\; &= \;\;
\frac{1}{\sqrt{2}}(|RR\rangle\pm|LL\rangle)^{AB}_{P}, \\
|\psi^\pm\rangle^{AB}_{P} \;\; &= \;\;
\frac{1}{\sqrt{2}}(|RL\rangle\pm|LR\rangle)^{AB}_{P},
 \end{split}
\end{equation}
and the four Bell states in the spatial-mode DOF as
\begin{equation}                           \label{eq.7}
\begin{split}
|\phi^\pm\rangle^{AB}_{S} \;\; &= \;\; \frac{1}{\sqrt{2}}(|a_1b_1\rangle\pm|a_2b_2\rangle)^{AB}_{S}, \\
|\psi^\pm\rangle^{AB}_{S} \;\; &= \;\;
\frac{1}{\sqrt{2}}(|a_1b_2\rangle\pm|a_2b_1\rangle)^{AB}_{S}.
 \end{split}
\end{equation}
Also we refer to the states $|\psi^\pm\rangle^{AB}_{P}$ and
$|\psi^\pm\rangle^{AB}_{S}$ as the odd-parity states, and
$|\phi^\pm\rangle^{AB}_{P}$ and $|\phi^\pm\rangle^{AB}_{S}$ as the
even-parity states.

\begin{figure}[!h]
\centering\includegraphics[width=12 cm,angle=0]{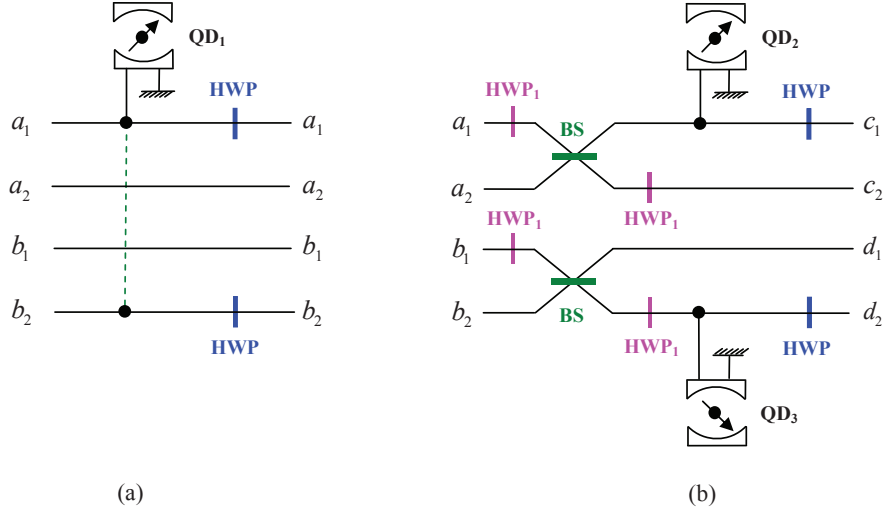}
\caption{Schematic  diagram  of  the  present  HBSA  protocol  for
the spatial-mode entangled Bell states, without destroying the
polarization Bell states of the photon pair $AB$. (a) The QND is
used to distinguish the odd-parity states $|\psi^\pm\rangle^{AB}_S$
from the even-parity states $|\phi^\pm\rangle^{AB}_S$. (b)  The QND
is used to distinguish the "+" phase state $|\psi^+\rangle^{AB}_S$
($|\phi^+\rangle^{AB}_S$) from the "-" phase states
$|\psi^-\rangle^{AB}_S$ ($|\phi^-\rangle^{AB}_S$). The dashed line
presents the case that the photons $A$ coming from the spatial mode
$\vert a_1\rangle$ and $B$ coming from $\vert b_2\rangle$ pass
through QD$_1$ in sequence. The small mirror is used to reflect the
photon for interacting with the cavity twice. HWP represents a
half-wave plate which is used to perform a phase-flip operation
$Z=|R\rangle\langle R|-|L\rangle\langle L|$ in the polarization DOF,
while HWP$_1$ represents another half-wave plate which is used to
perform a bit-flip operation $X=|R\rangle\langle L|+|L\rangle\langle
R|$ in the polarization DOF. BS represents a 50:50 beam splitter.}
\label{figure2}
\end{figure}

\subsection{HBSA protocol for Bell states in spatial-mode DOF}

\label{sec31}

The optical properties of a  singly  charged  QD in a
strong-coupling single-side microcavity,  which has been used for a
controlled-phase gate,  can be used to construct a quantum
nondemolition detector (QND), as shown in Fig.\ref{figure2}.   Let
us assume that the initial states of the excess electron in the
cavity and a single photon injected are
$(|\uparrow\rangle+|\downarrow\rangle)/\sqrt{2}$ and
$\alpha|R\rangle+\beta|L\rangle$, respectively. By adjusting the
frequencies $\omega-\omega_c\approx\kappa/2$ to get the phase shift
difference between the left and the right circular polarization
lights as $\Delta\varphi=\pi/2$, the function of a single photon
interacting with  a QD-cavity system twice is
\begin{eqnarray}                           \label{eq.8}
(\alpha|R\rangle+\beta|L\rangle) \otimes
\frac{1}{\sqrt{2}}(|\uparrow\rangle+|\downarrow\rangle)\;\;\rightarrow\;\;
e^{2i\varphi_0}(\alpha|R\rangle-\beta|L\rangle) \otimes
\frac{1}{\sqrt{2}}(|\uparrow\rangle-|\downarrow\rangle).
\end{eqnarray}
One can detect whether or not there is a photon interacting with the
QD-cavity system by measuring the spin state of the excess electron
with the orthogonal basis $\{|\pm\rangle=(|\uparrow\rangle\pm
|\downarrow\rangle)/\sqrt{2}\}$. If the excess electron is in the
state $|-\rangle$, there is a photon interacting with the QD-cavity
system (With a phase-flip operation $Z=|R\rangle\langle
R|-|L\rangle\langle L|$ on the photon, its original polarization
state is recovered).  Otherwise, there is no photon (or there are
two photons) interacting  with the QD-cavity system when the state
of excess electron doesn't change. Therefore, if the state of excess
electron in QD is not changed, there are an even number of photons
detected by the QD-cavity system. With this principle, the QD-cavity
system can be used as a QND to distinguish the case with an even
number of photons from that with an odd number of photons.

Now, we will use this photon-number QND to construct a parity-check
QND for the spatial-mode states of a photon pair. If the excess
electrons in QD$_1$ in Fig.\ref{figure2}(a) is  prepared initially
in the state $|+\rangle$, after the photons $A$ and $B$ pass through
QD$_1$ in sequence, the state of the excess electron is in
$|+\rangle$ with the input states $|\psi^\pm\rangle^{AB}_S$
(odd-parity). However, the state of the excess electron  becomes
$|-\rangle$ if the input states are $|\phi^\pm\rangle^{AB}_S$
(even-parity). By applying a Hadamard gate on the excess electron
spin, the spin superposition states $\vert +\rangle$ and $\vert
-\rangle$ can be rotated to the states $|\uparrow\rangle$ and
$|\downarrow\rangle$, respectively. If we have an auxiliary photon
$c$ in the initial state $\vert
\varphi_c\rangle=(|R\rangle+|L\rangle)/\sqrt{2}$ and let it pass
through QD$_1$, after it is reflected from the cavity, the state of
the system composed of the photon and the QD electron spin becomes
\begin{equation}                           \label{eq.9}
\begin{split}
\frac{1}{\sqrt{2}}(|R\rangle+|L\rangle)|\uparrow\rangle  \;\; &\rightarrow\;\;  \frac{1}{\sqrt{2}}e^{i\varphi_0}(|R\rangle+i|L\rangle)|\uparrow\rangle,\\
\frac{1}{\sqrt{2}}(|R\rangle+|L\rangle)|\downarrow\rangle \;\;
&\rightarrow\;\;
\frac{1}{\sqrt{2}}e^{i\varphi_0}(|R\rangle-i|L\rangle)|\downarrow\rangle.
 \end{split}
\end{equation}
The output state of the auxiliary photon $c$ can be measured in
orthogonal linear polarization basis. If the auxiliary photon $c$ is
in state $(|R\rangle+i|L\rangle)/\sqrt{2}$, the state of the excess
electron in QD is $|\uparrow\rangle$. Otherwise the state of the
excess electron in QD is $|\downarrow\rangle$. In this way, one can
construct a parity-check QND for the spatial-mode states to
distinguish the odd-parity states $|\psi^\pm\rangle^{AB}_S$ from the
even-parity states $|\phi^\pm\rangle^{AB}_S$ by detecting the spin
state of the excess electron in QD$_1$. The spin state of excess
electron in QD$_1$ is changed for even-parity states and unchanged
for odd-parity states.

With the QND in Fig.\ref{figure2}(a), the four Bell states in the
spatial-mode DOF are divided into two groups
$|\psi^\pm\rangle^{AB}_S$ and $|\phi^\pm\rangle^{AB}_S$.  The next
task of BSA in spatial-mode DOF is to distinguish the different
relative phases in each group. The QND shown in Fig.\ref{figure2}(b)
is used to distinguish the Bell states with the relative phase zero
from those with the relative phase $\pi$. BS can accomplish the
following transformations in spatial-mode DOF,
\begin{equation}                           \label{eq.10}
\begin{split}
|a_1\rangle \;\; &\rightarrow \;\;
\frac{1}{\sqrt2}(|c_1\rangle+|c_2\rangle),\\
|a_2\rangle \;\; &\rightarrow \;\;  \frac{1}{\sqrt2}(|c_1\rangle-|c_2\rangle), \\
|b_1\rangle \;\; &\rightarrow \;\;
\frac{1}{\sqrt2}(|d_1\rangle+|d_2\rangle), \\
|b_2\rangle \;\; &\rightarrow\;\;
\frac{1}{\sqrt2}(|d_1\rangle-|d_2\rangle).
 \end{split}
\end{equation}
After the  operations by BSs in  Fig.\ref{figure2}(b),  the two
groups of Bell states become:
\begin{equation}                           \label{eq.11}
\begin{split}
|\phi^+\rangle^{AB}_S \; = \;
\frac{1}{\sqrt2}(|a_1b_1\rangle+|a_2b_2\rangle)^{AB}_S
\;\;&\rightarrow\;\;
|\phi^+\rangle^{AB}_S \; = \; \frac{1}{\sqrt2}(|c_1d_1\rangle+|c_2d_2\rangle)^{AB}_S,  \\
|\phi^-\rangle^{AB}_S \; = \;
\frac{1}{\sqrt2}(|a_1b_1\rangle-|a_2b_2\rangle)^{AB}_S
\;\;&\rightarrow\;\;
|\psi^+\rangle^{AB}_S \; = \; \frac{1}{\sqrt2}(|c_1d_2\rangle+|c_2d_1\rangle)^{AB}_S, \\
|\psi^+\rangle^{AB}_S \; = \;
\frac{1}{\sqrt2}(|a_1b_2\rangle+|a_2b_1\rangle)^{AB}_S
\;\;&\rightarrow\;\;
|\phi^-\rangle^{AB}_S \; = \; \frac{1}{\sqrt2}(|c_1d_1\rangle-|c_2d_2\rangle)^{AB}_S, \\
|\psi^-\rangle^{AB}_S \; = \;
\frac{1}{\sqrt2}(|a_1b_2\rangle-|a_2b_1\rangle)^{AB}_S
\;\;&\rightarrow\;\; |\psi^-\rangle^{AB}_S \; = \;
\frac{1}{\sqrt2}(|c_1d_2\rangle-|c_2d_1\rangle)^{AB}_S.
 \end{split}
\end{equation}
That is,  $|\phi^+\rangle^{AB}_S$, $|\phi^-\rangle^{AB}_S$,
$|\psi^+\rangle^{AB}_S$, and $|\psi^-\rangle^{AB}_S$ become
$|\phi^+\rangle^{AB}_S$,  $|\psi^+\rangle^{AB}_S$,
$|\phi^-\rangle^{AB}_S$, and $|\psi^-\rangle^{AB}_S$, respectively.
With the parity-check measurement shown in Fig.\ref{figure2}(b), one
can read out the information about the relative phases in the groups
$|\phi^\pm\rangle^{AB}_S$ and $|\psi^\pm\rangle^{AB}_S$. If the
states of the excess electrons in QD$_2$ and QD$_3$ are both changed
(unchanged), the state input is $|\phi^-\rangle^{AB}_S$ or
$|\psi^-\rangle^{AB}_S$, and the output ports are $c_1$ and $d_2$
($c_2$ and $d_1$). While the state of the excess electron in QD$_2$
is changed (unchanged) and the state of the excess electron in
QD$_3$ is unchanged (changed), the state input is
$|\phi^+\rangle^{AB}_S$ or $|\psi^+\rangle^{AB}_S$, and the output
ports are $c_1$ and $d_1$ ($c_2$ and $d_2$).

The relation between the initial spatial-mode Bell states and the
outcomes of the QNDs is shown in Table \ref{table1}. The two-photon
system is in one of the two odd-parity states
$|\psi^{\pm}\rangle^{AB}_S$ in spatial-mode DOF if the state of
excess electron in QD$_1$ is unchanged. When the state of the excess
electron in QD$_1$ is changed, the two-photon system is in one of
the two even-parity states $|\phi^{\pm}\rangle^{AB}_S$. With QD$_2$
and QD$_3$, we read out the information about the relative phases in
the groups $|\phi^\pm\rangle^{AB}_S$ and $|\psi^\pm\rangle^{AB}_S$.
Therefore, for the state $|\psi^+\rangle^{AB}_S$, the state of the
excess electron in QD$_1$ is unchanged, and the states of the excess
electrons in QD$_2$ and QD$_3$ are in combination of one changed and
the other unchanged. If the state of  the excess electron in QD$_1$
is unchanged and the states of the excess electrons in QD$_2$ and
QD$_3$ are both changed or unchanged, the input state of the
two-photon system is $|\psi^-\rangle^{AB}_S$. For the two-photon
state $|\phi^+\rangle^{AB}_S$, the state of the excess electron in
QD$_1$ is changed, and the states of the excess electrons in QD$_2$
and QD$_3$ are in combination of one changed and the other
unchanged. For the input state $|\phi^-\rangle^{AB}_S$, the state of
the excess electron in QD$_1$ is changed, and the states of the
excess electrons in QD$_2$ and QD$_3$ are both changed or unchanged.

From the preceding analysis, one can see that the roles of the two
QNDs are accomplishing the task of parity check. The first QND can
distinguish the two even-parity states in spatial-mode DOF from the
two odd-parity states. With two BSs, the two states with two
different relative phases are transformed into another two states
with different parities. After the second QND, one can distinguish
the four Bell states in spatial -mode DOF without destroying the two
photons, which provides the convenience for the BSA in polarization
DOF.

\begin{table}[htb]
\centering \caption{Relation between the four Bell states in the
spatial-mode DOF and the output results of the measurements on
electron-spin states.}
\begin{tabular}{cccc}
\hline

\;\;\;\;\;\; Bell  $\;\;\;\;\;\;\;\;\;\;$    & \multicolumn {3}{c}{Results} \\
\cline{2-4}
    States               &          QD$_1$           &          QD$_2$       &        QD$_3$           \\
   \hline

 $|\psi^{+}\rangle^{AB}_S$    &       unchange        &         change (unchange)          &     unchange (change)    \\

 $|\psi^{-}\rangle^{AB}_S$    &       unchange        &         change (unchange)          &     change (unchange)      \\

 $|\phi^{+}\rangle^{AB}_S$    &       change          &         change (unchange)            &   unchange (change)       \\

 $|\phi^{-}\rangle^{AB}_S$    &       change          &         change (unchange)           &    change (unchange)      \\ \hline
\end{tabular}\label{table1}
\end{table}

\subsection{HBSA protocol for Bell states in polarization DOF}

Now let us move our attention to distinguish the four Bell states
$|\psi^\pm\rangle^{AB}_P$ and $|\phi^\pm\rangle^{AB}_P$ in
polarization. Hu et al. \cite{QD4} showed that the four Bell states
in polarization DOF can be easily distinguished with two photons
passing through a one-side QD-cavity system, and this nonlinear
optical effect of a one-side QD-cavity system can be used in our
HBSA protocol for polarization DOF, as the same as that by  Hu et
al. \cite{QD4}.

\begin{figure}[!h]
\centering
\includegraphics[width=6.6 cm,angle=0]{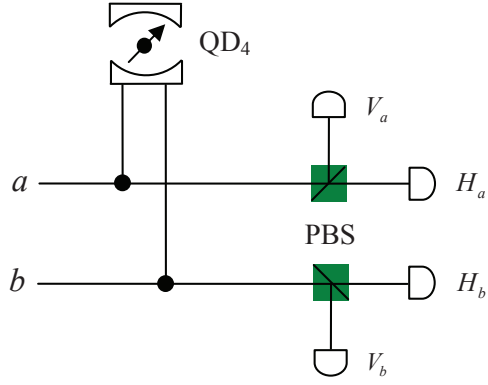}
\caption{Schematic  diagram  of  the  present  BSA  protocol  for
polarization  Bell states. The two spatial modes $a$ and $b$ are
sent into the cavity in sequence.} \label{figure3}
\end{figure}

Figure \ref{figure3} is the proposal for BSA in polarization DOF. If
we have the two spatial modes selected in Sec. \ref{sec31} put to
the cavity in sequence after the BSA in spatial-mode DOF and adjust
the frequencies $\omega-\omega_c\approx\kappa/2$ to get
$\Delta\varphi=\pi/2$, we can get the transformations as follows.
\begin{equation}                           \label{eq.12}
\begin{split}
\frac{1}{2}(|RR\rangle\pm|LL\rangle)\otimes(|\uparrow\rangle+|\downarrow\rangle)
\;\; & \rightarrow \;\;  \frac{1}{2}e^{2i\varphi_0}[(|RR\rangle
\mp|LL\rangle)\otimes(|\uparrow\rangle-|\downarrow\rangle)], \\
\frac{1}{2}(|RL\rangle\pm|LR\rangle)\otimes(|\uparrow\rangle+|\downarrow\rangle)
\;\; &\rightarrow \;\;
\frac{1}{2}e^{i(\varphi_0+\varphi_h)}[(|RL\rangle
\pm|LR\rangle)\otimes(|\uparrow\rangle+|\downarrow\rangle)].
 \end{split}
\end{equation}
If QD$_4$ in Fig. \ref{figure3} is prepared initially in the state
$|+\rangle$, after the interaction of two photons with the QD-cavity
system, one can identify whether the two-photon input states are the
Bell states $|\psi^\pm\rangle^{AB}_P$ (corresponding to spin
$|+\rangle$) or  $|\phi^\pm\rangle^{AB}_P$ (corresponding to spin
$|-\rangle$) by measuring the excess electron-spin state. Measuring
the two photons in the polarization basis $\{|H\rangle,
|V\rangle\}$, it is possible to distinguish $|\psi^+\rangle^{AB}_P$
and $|\phi^+\rangle^{AB}_P$ from $|\psi^-\rangle^{AB}_P$ and
$|\phi^-\rangle^{AB}_P$, respectively. The relation between the
initial Bell states in the polarization DOF and the results of the
measurements on QD$_4$ and the two photons with the basis
$\{|H\rangle, |V\rangle\}$ is shown in Table \ref{tab2}.

\begin{table}
\centering\caption {The relation between the initial Bell states in
the polarization DOF and the output results of the QD$_4$ and the
single-photon detections.}
\begin{tabular}{ccc}
\hline
 Bell     & \multicolumn {2}{c}{Results} \\ \cline{2-3}
    States                 &           QD$_4$         &       Detector                \\
   \hline

  $|\psi^{+}\rangle^{AB}_P$    &         unchange       &  $\{H_a, H_b\}$  or $\{V_a, V_b\}$          \\

  $|\psi^{-}\rangle^{AB}_P$    &         unchange       &  $\{H_a, V_b\}$  or $\{V_a, H_b\}$               \\

  $|\phi^{+}\rangle^{AB}_P$    &         change         &  $\{H_a, V_b\}$  or $\{V_a, H_b\}$             \\

  $|\phi^{-}\rangle^{AB}_P$    &         change         &  $\{H_a, H_b\}$  or $\{V_a, V_b\}$              \\ \hline
\end{tabular} \label{tab2}
\end{table}

By far,  we have described the principle of our complete and
deterministic HBSA with the nonlinear optics in one-side QD-cavity
systems. The BSA on the spatial and the polarization-mode DOFs can
be realized by adjusting the frequencies
$\omega-\omega_c\approx\kappa/2$ to get the  phase shift
$\Delta\varphi=\pi/2$.

\section{Applications of HBSA in quantum communication}

As a complete and deterministic analysis on quantum states is
important in quantum communication, it is interesting to discuss the
applications of HBSA. Let us use hyperentanglement swapping as an
example to describe its principle.

\begin{figure}[!h]
\centering
\includegraphics[width=11 cm,angle=0]{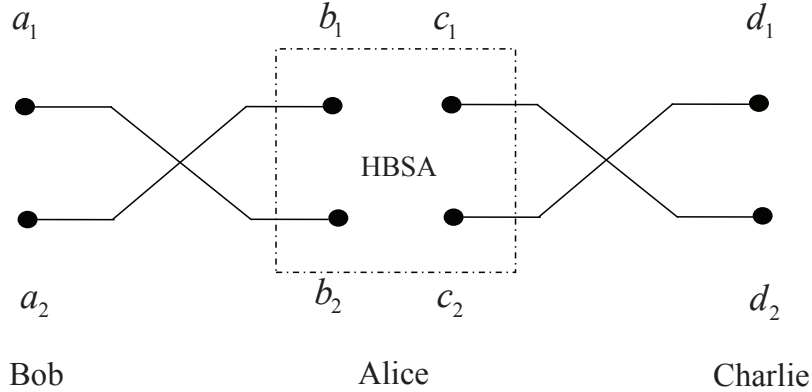}
\caption{Schematic diagram for the hyperentanglement swapping in
both the polarization and the spatial-mode DOFs. The initial
hyperentangled states are prepared in nodes $AB$ and $CD$ (also the
four photons). After Alice performs the HBSA on the two photons
$BC$, Bob and Charlie can get the hyperentangled state between nodes
$A$ and $D$.} \label{figure4}
\end{figure}

Hyperentanglement swapping  enables two parties in quantum
communication to obtain hyperentanglement between two particles
which do not interact with each other initially. Suppose that the
two entangled pairs $AB$ and $CD$ are in the following
hyperentangled states:
\begin{equation}                           \label{eq.13}
\begin{split}
|\Phi^+\rangle^{AB}_{PS} \;\; &= \;\; \frac{1}{2}(|RR\rangle+|LL\rangle)^{AB}_P\otimes(|a_1b_1\rangle+|a_2b_2\rangle)^{AB}_S, \\
|\Phi^+\rangle^{CD}_{PS} \;\; &= \;\;
\frac{1}{2}(|RR\rangle+|LL\rangle)^{CD}_P\otimes(|c_1d_1\rangle+|c_2d_2\rangle)^{CD}_S.
 \end{split}
\end{equation}
The superscripts $A$ and $B$ denote that the particles are in nodes
$A$ and $B$, respectively, as shown in Fig. \ref{figure4}. Alice
shares a photon pair $AB$ with Bob, and she also shares a photon
pair $CD$ with Charlie. The task of this hyperentanglement-swapping
protocol is to entangle the two photons $A$ and $D$ in both the
polarization and the spatial-mode DOFs.

To complete entanglement swapping of hyperentangled states, Alice
performs HBSA on the two particles $B$ and $C$ in her hand, as shown
in Fig. \ref{figure4}. The state of the whole system can be
rewritten as
\begin{equation}                           \label{eq.14}
\begin{split}
|\Phi^+\rangle^{AB}_{PS}\otimes|\Phi^+\rangle^{CD}_{PS}\;\; &= \;\;
\frac{1}{4}[(|\phi^+\rangle^{AD}_P|\phi^+\rangle^{BC}_P+|\phi^-\rangle^{AD}_P
|\phi^-\rangle^{BC}_P+|\psi^+\rangle^{AD}_P|\psi^+\rangle^{BC}_P  \\
\;\; & + \;\; |\psi^-\rangle^{AD}_P|\psi^-\rangle^{BC}_P)
\otimes(|\phi^+\rangle^{AD}_S|\phi^+\rangle^{BC}_S+|\phi^-\rangle^{AD}_S|\phi^-\rangle^{BC}_S\\
\;\; & +\;\; |\psi^+\rangle^{AD}_S
|\psi^+\rangle^{BC}_S+|\psi^-\rangle^{AD}_S|\psi^-\rangle^{BC}_S)].
 \end{split}
\end{equation}
If the outcome of HBSA is
$|\phi^+\rangle^{BC}_P|\phi^+\rangle^{BC}_S$, the two photons
located in the nodes $A$ (Bob) and $D$ (Charlie) is in the
hyperentangled state $|\phi^+\rangle^{AD}_P|\phi^+\rangle^{AD}_S$.
The other outcomes lead to the other hyperentangled states, such as
$|\phi^+\rangle^{AD}_P|\phi^-\rangle^{AD}_S$,
$|\phi^-\rangle^{AD}_P|\phi^\pm\rangle^{AD}_S$,
$|\phi^\pm\rangle^{AD}_P|\psi^\pm\rangle^{AD}_S$,
$|\psi^\pm\rangle^{AD}_P|\phi^\pm\rangle^{AD}_S$, and
$|\psi^\pm\rangle^{AD}_P|\psi^\pm\rangle^{AD}_S$. In principle, it
is not difficult for Bob and Charlie to transform their
hyperentangled state into the form
$|\phi^+\rangle^{AD}_P|\phi^+\rangle^{AD}_S$. For instance, if Bob
and Charlie obtain the state
$|\psi^-\rangle^{AD}_P|\psi^-\rangle^{AD}_S$, the state
$|\phi^+\rangle^{AD}_P|\phi^+\rangle^{AD}_S$ can be obtained in the
way that Charlie performs an operation $-\sigma_y=|R\rangle\langle
L|-|L\rangle\langle R|$ in polarization (both the two spatial modes
 $d_1$ and $d_2$) and then exchanges the two
spatial modes after he introduces a phase $\pi$ in the spatial mode
$d_1$ with a $\lambda/2$ wave plate.

The hyperentanglement-swapping protocol presented here is completed
by two simultaneous but independent processes, including both the
polarization BSA and the spatial-mode BSA. If we only perform the
Bell-state measurement on the photons $B$ and $C$ in the
polarization DOF, photons $A$ and $D$ will be entangled in the
polarization degree of freedom but leave their state in the
spatial-mode DOF be a mixed one.

\section{Discussion and conclusion}

BSA is essential in quantum communication, especially in
long-distance quantum communication assisted by quantum repeater.
There are many proposals for analyzing Bell states in polarization
photon pairs. For the hyperentangled BSA discussed here, the
entanglements in different DOFs need to be analyzed independently.
This is different from the hyperentanglement-assisted BSA in
polarization DOF, in which another degree of freedom is used as an
additional system and is consumed in the analysis.

In our proposal, the BSA in the polarization and the spatial-mode
DOFs are completed by the relative phase shift $\pi/2$ of left
circularly and right circularly polarized lights. In 2011, Young et
al. \cite{con} performed high-resolution reflection spectroscopy of
a quantum dot resonantly coupled to a pillar microcavity with
quality factor $Q\sim51000$ ($d=2.5\mu$m), and their experiment
showed that a QD-induced phase shift of 0.2 rad between an
(effectively) empty cavity and a cavity with a resonantly coupled QD
can be deduced, by using a single-photon level probe. If the cavity
loss is $\kappa_s<1.3\kappa$ \cite{QD4}, by improving the mode
matching and the pillar design, this relative phase shift could
achieve $\pm\pi/2$. In an ideal condition, the fidelity of the HBSA
proposal can reach 100\%, but in experiment the fidelity is reduced
by the limitation of techniques. When two photons are put into a
cavity in sequence, the time interval $\Delta t$ between the two
photons should be shorter than the electron spin decoherence time
$T^e_2$. The electron spin decoherence time $T^e_2$ could be
extended to $\mu$s using spin echo techniques, which is longer than
the time interval $\Delta t$ ($n$s) with weak excitation \cite{QD4}.
The heavy-light hole mixing can reduce the fidelity \cite{hole1},
but it can be improved for charged excitons due to the quenched
exchanged interaction \cite{hole2, hole3}. The trion dephasing
effect can also reduce the fidelity \cite{trion1, trion2, trion3},
but this dephasing effect of $X^-$ can be neglected with the hole
spin coherence time three orders of magnitude longer than the cavity
photon lifetime \cite{trion4, trion5, trion6}. As the decoherence
effects of electron spin, heavy-light hole, and $X^-$ can be
neglected, the main factors that reduce the fidelity of HBSA
proposal are the coupling strength and the cavity side leakage.

\begin{figure}                
\begin{minipage}[t]{0.5\linewidth}
\centering
\includegraphics[width=2.4in]{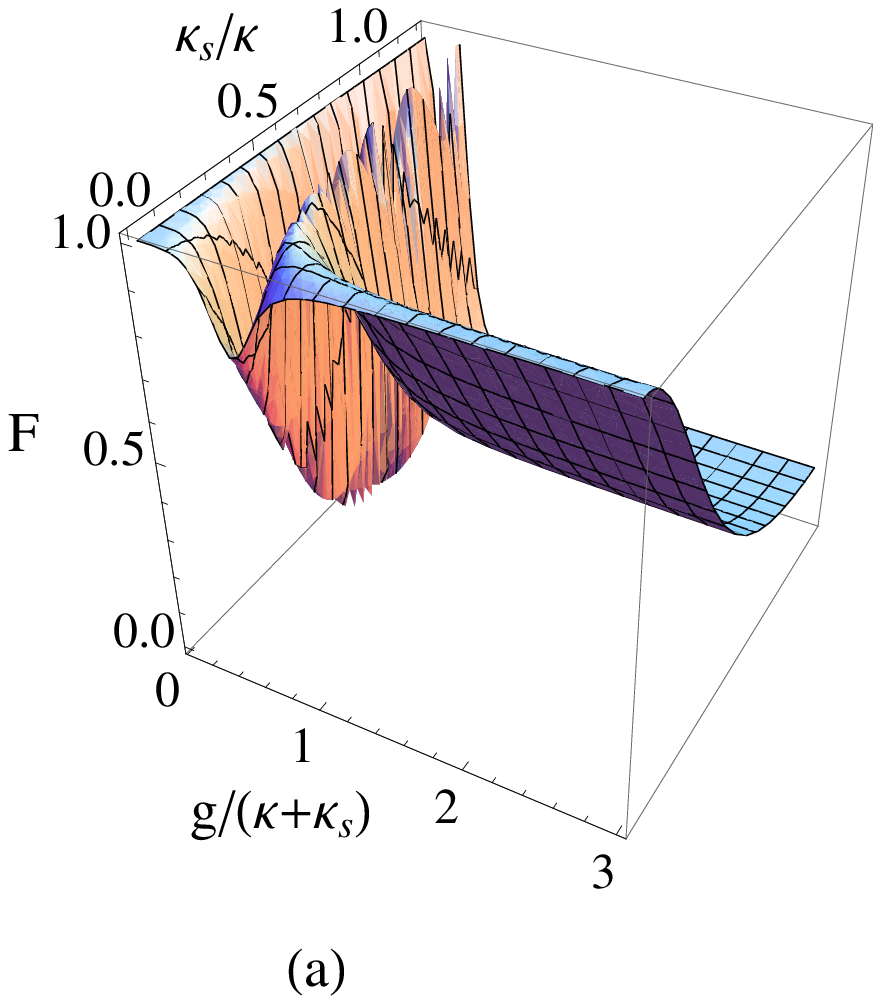}
\end{minipage}%
\begin{minipage}[t]{0.5\linewidth}
\centering
\includegraphics[width=2.4in]{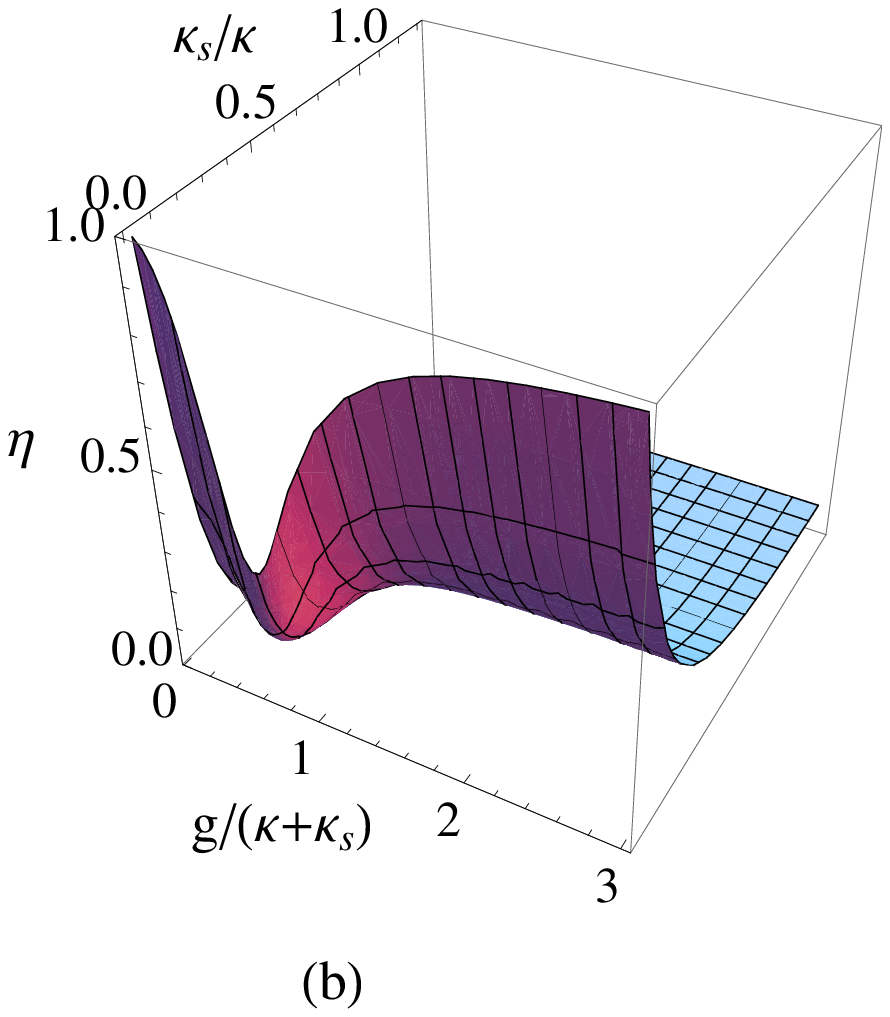}
\end{minipage}
\caption{ The fidelity (a) and the efficiency (b) of the present
HBSA protocol for the hyperentangled-Bell state
$|\phi^+\rangle_P|\phi^+\rangle_S$ vs the coupling strength
$g/(\kappa+\kappa_s)$ and the side leakage rate $\kappa_s/\kappa$
with $\gamma=0.1\kappa$.} \label{figure5}
\end{figure}

If the cavity side leakage is neglected, the fidelity of the HBSA
proposal can reach 100\% in the strong-coupling regime with
$|r_0(\omega)|\cong1$ and $|r_h(\omega)|\cong1$. However, there is
rigorous limitation in the QD-micropillar cavity, and the cavity
side leakage should be considered. Defining fidelity as
$F=|\langle\psi_f|\psi\rangle|^2$, the fidelities for HBSA proposal
can be calculated. Here $|\psi_f\rangle$ is the final state of the
total system which includes the external reservoirs, and
$|\psi\rangle$ is the final state with an ideal condition. As
discussed in Ref. \cite{QD4}, the fidelity of even parity Bell
states is larger than odd parity Bell states in polarization DOF, we
calculate the fidelity (in amplitude) and the efficiency of the
present HBSA proposal for the hyperentangled state
$|\phi^+\rangle_P|\phi^+\rangle_S$. The fidelity of this state is
\begin{eqnarray}                           \label{eq.15}
F&=&\frac{[(\zeta^{5}+\xi^{5})^2+22\epsilon^4(\zeta+\xi)^2+4\epsilon(\zeta^4-\xi^4)^2+16\epsilon^3(\zeta^2-\xi^2)^2+
9\epsilon^2(\zeta^3+\xi^3)^2]^2}
{(\zeta^{10}+\xi^{10})^2+22\epsilon^8(\zeta^2+\xi^2)^2+4\epsilon^2(\zeta^8-\xi^8)^2+16\epsilon^6(\zeta^4-\xi^4)^2+
9\epsilon^4(\zeta^6+\xi^6)^2} \times\frac{1}{128},
\end{eqnarray}
and the efficiency of this state is
\begin{eqnarray}                           \label{eq.16}
\eta=(\frac{1}{2}\zeta^4+\frac{1}{2}\xi^4)^2(\frac{1}{2}\zeta^2+\frac{1}{2}\xi^2)^2,
\end{eqnarray}
where $\zeta=|r_0|$, $\xi=|r_h|$, and $\epsilon=|r_0||r_h|$.
Figure.\ref{figure5} shows the fidelity and the efficiency of the
present HBSA protocol for the hyperentangled state
$|\phi^+\rangle_P|\phi^+\rangle_S$. From Fig.\ref{figure5}(a), we
can see that the HBSA proposal  works with a high fidelity in both
strong the coupling regime ($g>(\kappa+\kappa_s)/4$) and the weak
coupling regime ($g<(\kappa+\kappa_s)/4$). However, considering the
efficiency in Fig.\ref{figure5}(b), our protocol   only works
effectively in the strong coupling regime. In experiment, it is easy
to achieve the weak coupling, but it is challenging to observe a
strong coupling in a QD-cavity system. It has been observed the
coupling strength can be increased from $g\cong0.5(\kappa+\kappa_s)$
($Q=8800$) \cite{couple} to $g\cong2.4(\kappa+\kappa_s)$
($Q\sim40000$) \cite{couple1} for $d=1.5\mu$m  micropillar
microcavities by improving the sample designs, growth, and
fabrication \cite{couple2}. For the strong coupling
$g\cong0.5(\kappa+\kappa_s)$, the fidelity and efficiency are
$F=86\%$ and $\eta=36.8\%$ when $\kappa_s/\kappa=0$, and they are
$F=66\%$ and $\eta=0.7\%$ when $\kappa_s/\kappa=0.3$. For the strong
coupling $g\cong2.4(\kappa+\kappa_s)$, the fidelity and the
efficiency are $F=100\%$ and $\eta=95\%$ when $\kappa_s/\kappa=0$,
and they are $F=48\%$ and $\eta=15.1\%$ when $\kappa_s/\kappa=0.5$.
The quality factors in these micropillars are dominated by the side
leakage and cavity loss rate, so the top mirrors of  high-Q
micropillars ($d=1.5\mu$m) are thin down to get
$\kappa_s/\kappa\sim0.7$ and $g\cong\kappa+\kappa_s$  ($Q\sim17000$)
in Ref. \cite{QD4}. For the strong coupling $g\cong\kappa+\kappa_s$,
the fidelity and the efficiency are $F=97\%$ and $\eta=64\%$ when
$\kappa_s/\kappa=0$, but they are $F=33\%$ and $\eta=8.5\%$ when
$\kappa_s/\kappa\sim0.7$. Both the fidelity and the efficiency are
largely reduced by the cavity side leakage. To get small
$\kappa_s/\kappa$ in the strong coupling regime, high-efficiency
operation is highly demanded. This could be quite challenging for
micropillar microcavities. The recent experiments achieve the strong
coupling with large micropillars \cite{couple3}, while the side
leakage is small with small micropillars.

In summary, we have proposed a complete HBSA scheme with the
interaction between a circular polarization light and a one-side
QD-cavity system (the nonlinear optics of a one-side QD-cavity
system). We use the relative phase shift of the right and the left
circularly polarized lights to construct parity-check measurements
and analyze Bell states in different DOFs of two-photon systems. We
have also discussed its applications in long-distance quantum
communication processes in two different DOFs simultaneously.

\section*{ACKNOWLEDGEMENTS}

This work is supported by the National Natural Science Foundation of
China under Grant Nos. 10974020 and 11174039,  NCET-11-0031, and the
Fundamental Research Funds for the Central Universities.


\end{document}